\documentclass[a4paper,11pt]{article}
\pdfoutput=1 % if your are submitting a pdflatex (i.e. if you have
\usepackage{jinstpub} % for details on the use of the package, please
\title{\boldmath  Nonlinearity and pixel shifting effects in HXRG infrared detectors}

\author[a,1]{A. A. Plazas \note{Corresponding author.}}
\author[a]{C. Shapiro,}
\author[b]{R. Smith,}
\author[a,c,d]{J. Rhodes,}
\author[a]{E. Huff}

\affiliation[a]{Jet Propulsion Laboratory, California Institute of Technology,\\4800 Oak Grove Dr., Pasadena, CA 91109, USA}
\affiliation[b]{Caltech Optical Observatories, California Institute of Technology,\\1200 E. California Blvd., CA 91125, USA}
\affiliation[c]{California Institute of Technology,\\1200 E. California Blvd., CA 91125, USA}
\affiliation[d]{Institute for the Physics and Mathematics of the Universe,\\ 5-1-5 Kashiwanoha, Kashiwa, Chiba Prefecture 277-8583, Japan}

\emailAdd{andres.a.plazas.malagon@jpl.nasa.gov}

\abstract{We study the nonlinearity (NL) in the conversion from charge to voltage in infrared detectors (HXRG) for use in precision astronomy. We present laboratory measurements of the NL function of a H2RG detector and discuss the accuracy to which it would need to be calibrated in future space missions to perform cosmological measurements through the weak gravitational lensing technique. In addition, we present an analysis of archival data from the infrared H1RG detector of the Wide Field Camera 3 in the Hubble Space Telescope that provides evidence consistent with the existence of a sensor effect analogous to the ``brighter-fatter'' effect found in Charge-Coupled Devices. We propose a model in which this effect could be understood as shifts in the effective pixel boundaries, and discuss prospects of laboratory measurements to fully characterize this effect.}


\begin{document}
\maketitle
\flushbottom

\section{Introduction}
\label{sec:intro}
Weak gravitational lensing (WL)---the subtle distortion of the apparent shapes of background galaxies---is a powerful method to probe the mass distribution of the Universe and its evolution. The signal is small (the distortion of shapes by the large scale structure is of the order or $1$\textendash $2\%$) and current and future projects require that galaxy shapes be measured with sub-percent accuracy to avoid biases in the estimation of cosmological parameters from WL observations. The main source of noise is the intrinsic shape of the galaxies; however, it can be statistically reduced by averaging over a large number of galaxies, as is already being achieved by on-going projects (hundreds of millions of galaxies).  Therefore, the thorough understanding and mitigation of systematic errors\textemdash which can arise from numerous sources such telescope optics, the atmosphere, Point Spread Function (PSF) correction, shape measurement algorithms, detectors, etc.\textemdash becomes crucial for WL, in which the allowed measurement errors are of the order of $10^{-3}$ and $10^{-4}$ in size and shape, respectively (~\cite{a,b,c,d}).

Future space missions such as NASA's Wide Field Infrared Survey Telescope (WFIRST) will use WL as one of their main techniques to map large-scale structures and probe the dark Universe. The WL survey will be carried out by instruments with a new generation of hybrid near infrared (NIR) detectors manufactured by Teledyne Imaging Systems and known as Hawaii\footnote{HgCdTe Astronomical Wide Area Infrared Imager.}-XRG (HXRG), where X denotes the detector width in thousands of pixels. This family of detectors has been used by instruments such as NICMOS and WFC3 on board of the Hubble Space Telescope; however, they have not yet been used to perform shape measurements at the required accuracy for space-based WL. As with Charge-Coupled Devices (CCDs)\textemdash that have been used extensively in the past decades for optical astronomy\textemdash there are numerous detector effects that must be calibrated and corrected to satisfy the requirements to perform precision astronomy. In this paper we study two types of nonlinearities in NIR detectors: classical nonlinearity (NL) and the ``brighter-fatter" (BF) effect. The former refers to the deviation from linearity in the conversion of charge to voltage (and for this reason it is also known as voltage nonlinearity), while the latter---observed and characterized in CCDs and expected to be present in NIR detectors as well---is an effect in which the size of a point source increases with flux as charge accumulates. 

Section~\ref{sec:nir} presents a brief description of NIR detectors and highlights some of their features and differences with respect to CCDs. In Section~\ref{sec:nl} we show examples of laboratory measurements of classical nonlinearity and discuss its characterization and calibration processes, with weak lensing science in particular in mind. In Section~\ref{sec:bf} we present preliminary evidence for a BF-type effect in NIR detectors using existing archival data from the NIR detector of the Wide Field Camera 3 (WFC3-IR) onboard the Hubble Space Telescope (HST). We also discuss how these measurements could be understood in terms of a physical model that effectively shifts the pixel boundaries, and we discuss how laboratory measurements can help better characterize this effect. Finally, we conclude in Section~\ref{sec:end}.

\section{Near infrared detectors}
\label{sec:nir}
\begin{figure}[htbp]
\centering 
\includegraphics[width=0.5\textwidth,origin=c]{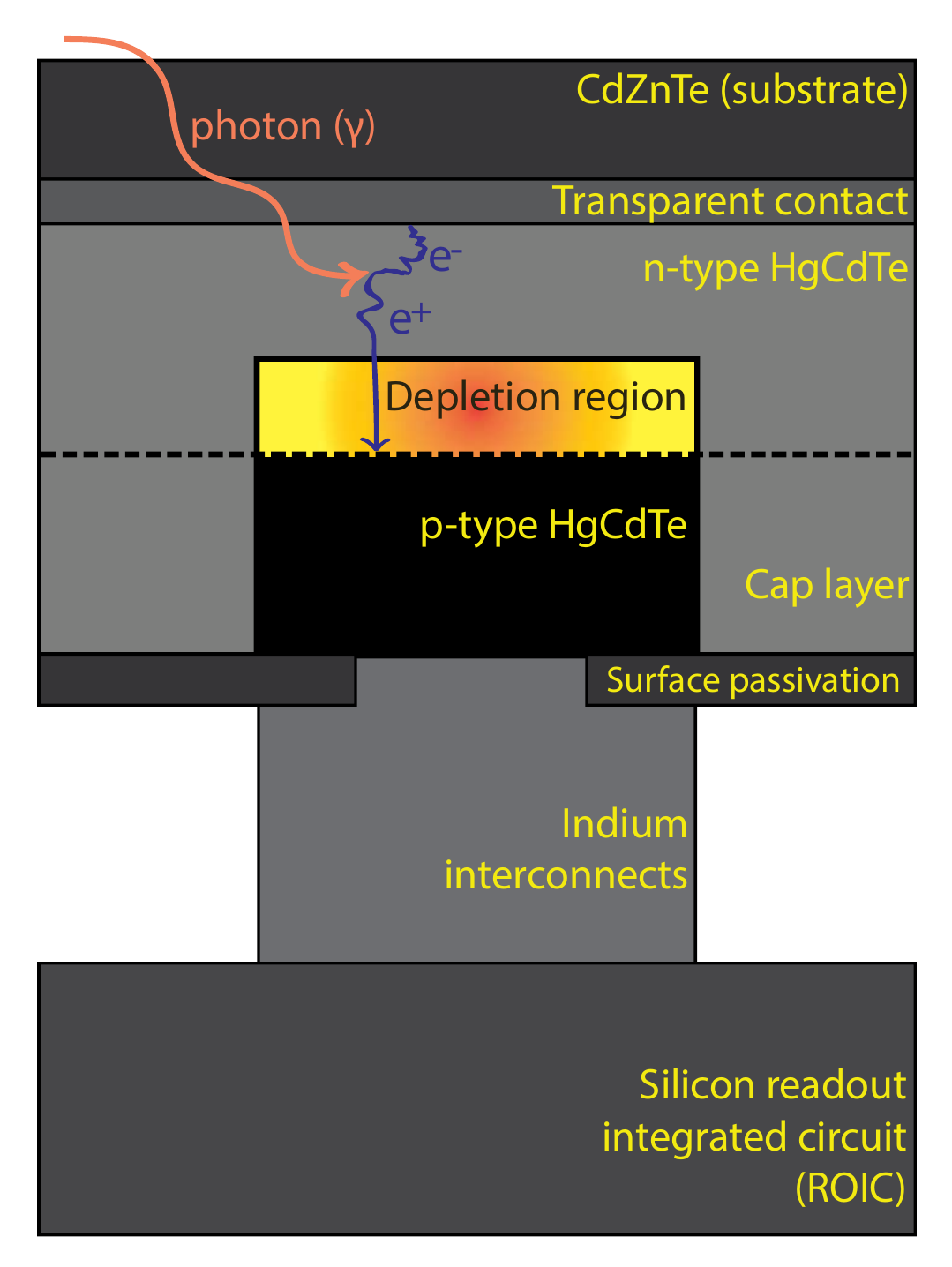}
\caption{\label{fig:nir} Diagram of a HgCdTe photodiode in a NIR sensor. An electron-hole pair is generated when a photon hits the n-type HgCaTe detection layer. The hole e$^+$ (in the case of a n-type material) diffuses to depletion region at the interface between the p- and n-type materials, where it is collected. The ROIC is connected to the HgCdTe layer through indium bumps. The CdZnTe substrate is used to grow the array of photodiodes and then removed. Adapted from Figure 1 of~\cite{e} and Figure 2 of~\cite{f}.}
\end{figure}

%. NIR observations also provide a window through the dust and gas surrounding nearby star-forming regions.
%For SNe observations, the WFC3/IR allows wide-area searches for high-redshift SNe, whose peak fluxes have been redshifted from optical to NIR wavelengths. Future space telescopes such as the Wide-Field Infrared Survey Telescope (WFIRST) will use infrared detectors and perform galaxy shape, photometric, and astrometric measurements for WL, SNe, and microlensing.
The NIR detectors that will be used by future space missions such as WFIRST consist of hybrid complementary metal-oxide semiconductors (CMOS) arrays, in which the process of photon collection is separated from signal readout (see Figure ~\ref{fig:nir}).
For a given pixel, photons are absorbed in a layer of HgCdTe (mercury cadmium telluride) through the photoelectric effect, producing a hole-electron pair. The hole then diffuses until it is collected in the depletion region generated at a p-n junction at the detector layer. This layer is connected to a readout integrated circuit (ROIC) through indium bumps. The voltage change induced by the photo-generated charge is then transferred to an analog switch matrix which supports sequential readout. The ROIC transmits the signal for all pixels to off-chip electronics, where it is digitized (\cite{e,f}). Note that, contrary to CCDs, the \emph{charge} is not transferred from pixel to pixel through vertical and serial registers. 

The sampling of the p-n diode voltage is nondestructive. If the sampling is performed once at the start and again at the end of an exposure, we obtain a correlated double sampling reading, similar to sampling the sense node of a CCD. However, more than two nondestructive samples are possible through Fowler sampling or Sampling Up the Ramp (SUTR), allowing the reduction of noise and the integration through cosmic rays hits. Despite the advantages of multiple nondestructive reading, other challenges arise, such as the electric coupling between pixels, known as inter-pixel capacitance (IPC), that has been shown to directly impact WL measurements (\cite{g}).  

NIR detectors have been used in HST and Spitzer instruments; and will be used in future space observatories such as the James Webb Space Telescope, WFIRST, and Euclid. NIR observations are a significant improvement for WL since they enable galaxy shape measurements at higher redshifts ($z\geq1$), which tightens constraints on dark energy and dark matter parameters. 

As an example, the Wide Field Imager of WFIRST will use 18 $4k\times4k$ HgCdTe NIR detectors, arranged in a 6$\times$3 layout and with a pixel size and scale of $10\ \mu$m and $0.11$ arc seconds per pixel, respectively. %The HgCdTe NIR detectors are manufactured by Teledyne Imaging Systems, and are part of a family of detectors known as Hawaii-XRG (HXRG), where X denotes the detector width in thousands of pixels. 

\section{Classical Nonlinearity}
\label{sec:nl}
\begin{figure}[htbp]
\centering 
\includegraphics[width=0.6\textwidth,origin=c]{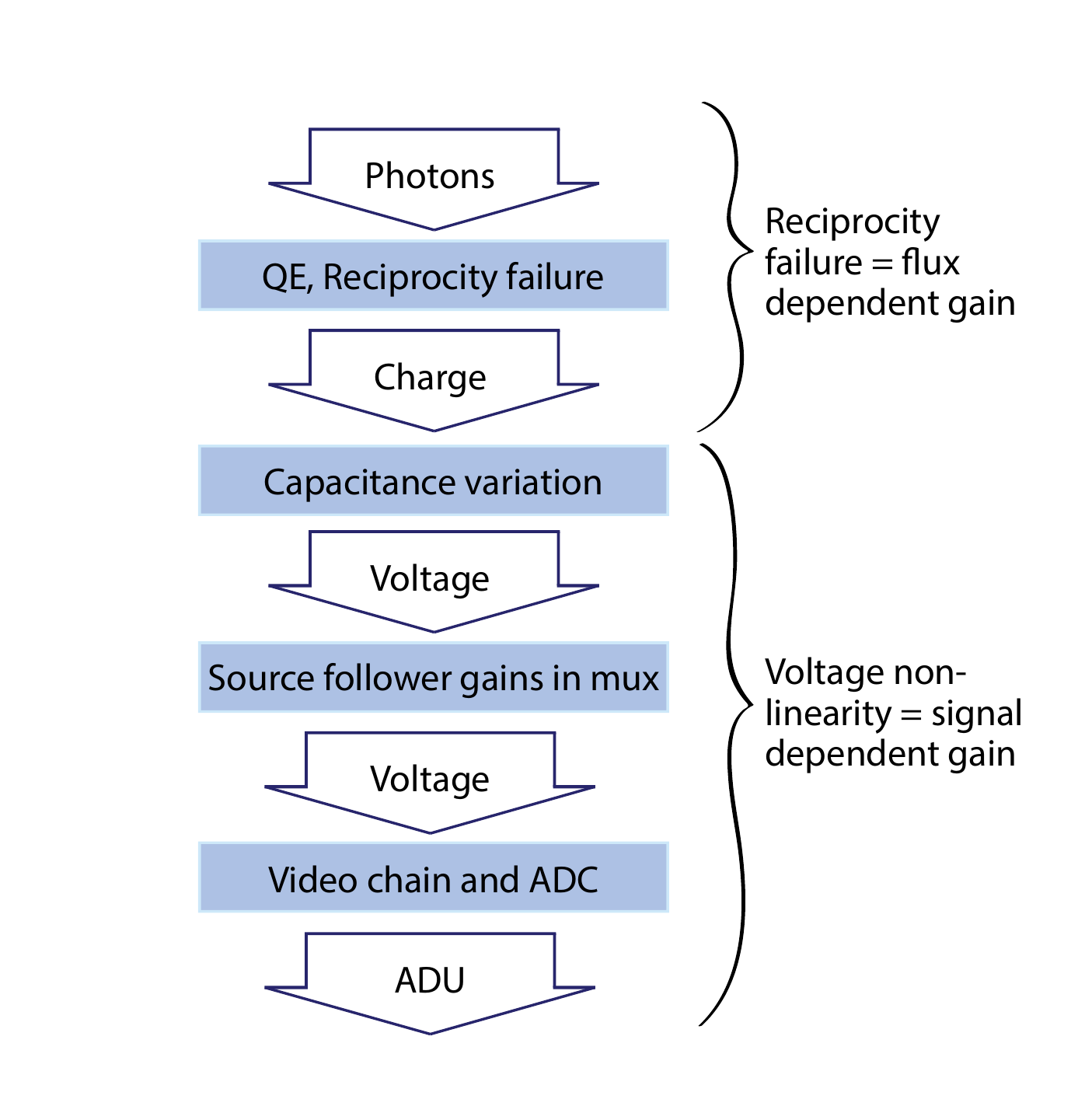}
\caption{\label{fig:nl} Total nonlinearity can be decomposed into independent components which occur in successive stages of the signal chain. These may be combined into a flux-dependent component presumed to occur in the photodiodes prior to conversion of charge to voltage (reciprocity failure), and a component which is dependent on integrated signal and includes charge-to-voltage conversion, subsequent gain stages, and analog-to-digital conversion (classical nonlinearity).}
\end{figure}

There are several steps in the signal chain where the conversion of photo-generated charge to voltage (or Analog-to-Digital Units, ADU) deviates from the ideal linear behavior (see Figure~\ref{fig:nl}).  
At low fluxes, the type on nonlinearity that dominates is known as count-rate nonlinearity or \emph{reciprocity failure} (as an analog to film photography)---the change in charge accumulation as a function of photon rate. As the flux increases, the relevant type of nonlinearity is in the charge to voltage conversion, and it is a consequence of the increasing junction capacitance as the depletion region narrows due to charge accumulation: the p-n diode acts as a parallel plate capacitor, and the effective junction capacitance is given by (\cite{h}):
\begin{equation}
\label{eq:cjn}
C_{\text{jn}}=A(1 + V/V_{\text{bi}})^{-1/2},
\end{equation}
where $A$ is the capacitance in the limit of large reverse bias, $V_{\text{bi}}$ is the ``built in" potential of the junction, and $V$ contains the change in potential due to accumulated charge (in addition to the difference between the constant potential at the diode cathode and the initial reset potential). Finally, nonlinearity can also arise due to the elements in the electronic chain, such as the source follower in the ROIC and the analog-to-digital conversion. Each of the previous types of nonlinearity are assumed to be independent. We consider the latter two in this analysis, and compose them into a single transfer function referred to as \emph{classical nonlinearity} (NL). 

In practice, when performing an astronomical observation, both RF and NL affect together the data through a single nonlinearity function. However, NL (and RF) are usually measured and calibrated by using laboratory data; they can be independently measured by illuminating the detector with a constant flux and recording the signal as a function of time (in the case of NL), or by applying a series of independently derived fluxes, while adjusting exposure time to keep signal nominally constant to maintain sign-to-noise-ratio and suppress NL (for RF). 

\subsection{NL measurements at the Precision Projector Laboratory}
We have carried out NL tests at JPL's Precision Projector Laboratory (PPL,~\cite{i,j})---a joint facility between NASA Jet Propulsion Laboratory and Caltech Optical Observatories that emulates realistic astronomical data by projecting sky scenes using optics which deliver high Strehl ratio over large format detectors. The chief design goal of the PPL is to emulate space-based astronomical observations to check the impact of detector effects on WL shape measurements. The emulation system is stable and versatile, with a wide range of control of the flux, wavelength, position, and f-number of the projected images. Images such as a grid of point sources can be focused over the entire area of a detector, which provides a multiplexed advantage over e.g. scanning a single bright spot.  The PPL data analysis pipeline also includes image processing and shape measurement tools. The lab has two H2RG detectors with have the same pixel scale ($18\ \mu$m) and wavelength cutoff ($1.7\ \mu$m) as the H1RG on the WFC3.
\begin{figure}[htbp]
\centering 
\includegraphics[width=1.0\textwidth,origin=c, page=1]{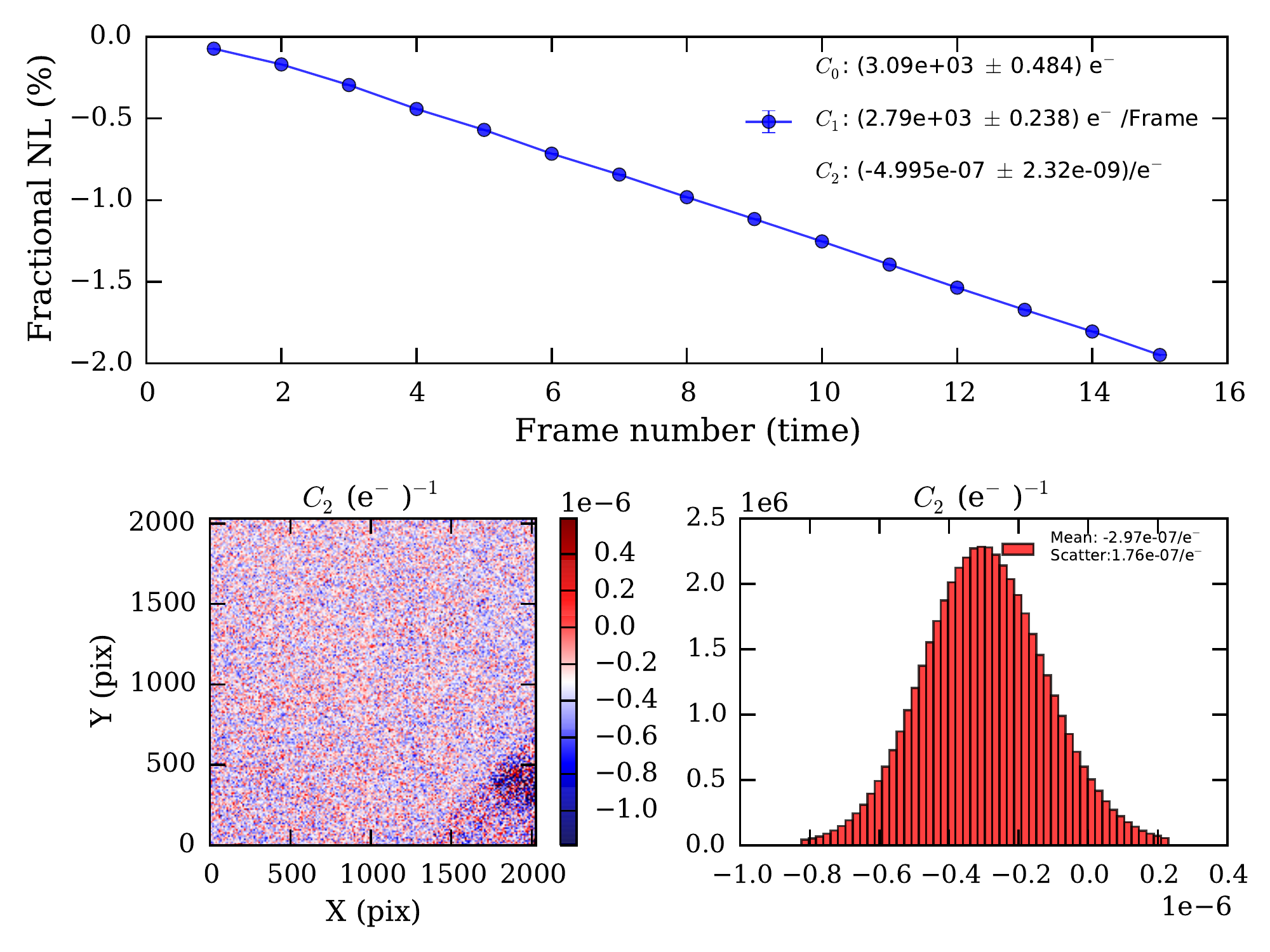}
\caption{\label{fig:nl_ppl} Example of NL measurements on a H2RG device at JPL's Precision Projector Laboratory. \emph{Upper panel}: Fractional NL obtained from fitting Equation~\ref{eq:fit} to the measured mean nonlinearity curve over all pixels in the PPL H2RG sensor (after 3\textendash$\sigma$ clipping rejection) and subtracting the linear component of the fit. The parameter $C_2$ can be directly identified with the parameter $\beta$ of Equation~\ref{eq:quadratic}. \emph{Lower left panel}: Map of the NL parameter coefficient $C_2$ along the whole detector. \emph{Lower right panel}: Distribution of $C_2$ after 3\textendash$\sigma$ clipping rejection. The conversion gain from ADU to elemental charge was measured to be approximately $2.7$ e/ADU.}
\end{figure}
Figure~\ref{fig:nl_ppl} shows results of the NL function measured from flat field images (after darks subtraction) of our H2RG detector. The measured curve (per pixel and averaged over all pixels) is usually approximated by a polynomial function. For low to medium signal (in voltage, ADU, or elemental charges) levels, it can be shown (see Appendix A) that Equation~\ref{eq:cjn}---in conjunction with the general relation between charge and voltage in a capacitor, $Q=CV$---leads to a quadratic correction to the sensed charge:
\begin{equation}
\label{eq:quadratic}
V(Q)=Q-\beta Q^2
\end{equation}
We fit a general quadratic polynomial, to allow for the off-set given by the reset voltage of the ramp, and a linear flux component of the signal ($C_1t$):
\begin{equation}
\label{eq:fit}
V=C_0 + C_1t + C_2(C_1t)^2
\end{equation}
Also, we have used the SUTR frame (or sample) number as a proxy of the time variable. The parameter $C_2$ can be directly identified with the parameter $\beta$ or Equation~\ref{eq:quadratic}.

In practice, the degree of the polynomial can be extended to higher orders. For instance, the NL correction performed in the WFC3-IR is derived from a third degree polynomial fit to a mean SUTR curve (\cite{p}). Depending on the particular science requirements, it must be investigated which degree provides a better fit to the data. Moreover, the question may arise of whether a polynomial approximation is correct given the form of Equation~\ref{eq:cjn}. A better approach may be to fit for the free parameters in Equation~\ref{eq:cjn} which is based on diode physics, assuming that nonlinearities in the electronics are negligible.

In addition, it must be kept in mind that different pixels could have different underlying linearity coefficients, and therefore it is also important to assess if it is consistent with the requirements at hand to accept a degree of mismatch between a given pixel and the mean behavior. The lower panels of Figure~\ref{fig:nl_ppl} show an example of the spatial variation that can be found in the determination of the NL model parameter $\beta$ ($C_2$). Thus, it is possible that a NL correction per (usable) pixel might be needed. However, it is usually challenging to measure NL for hundreds of millions of pixels with sufficient precision. Plazas et al. (2016) (\cite{k}) have studied the impact of NL on the WFIRST PSF for WL analysis using simulations. They find that, for a typical value of $\beta=5\times10^{-7}/$e$^-$, NL induces errors in the size and ellipticity of about $1\times10^{-2}$ and $2\times 10^{-3}$ respectively (in the H158 photometric band), larger than the required values of $1\times 10^{-3}$ and $4.7\times 10^{-4}$ on the knowledge of the size and ellipticity of the PSF in order not to bias cosmological parameter inferences from WL measurements by WFIRST. Plazas et al. (2016) also present fitting formulae to derive requirements on NL for the WFIRST detectors for different sets of tolerances on PSF properties. They also find that---for an example set of assumed requirements on PSF size and ellipticity and a quadratic model for NL---the NL model parameter $\beta$ should be calibrated to about $1\%$ to $2.4\%$ per pixel. 

\section{The ``brighter-fatter" effect}
\label{sec:bf}
Another nonlinear effect that has direct impact on PSF estimation and shape measurement for WL science is known as the ``brighter-fatter" (BF) effect (\cite{l}), in which point sources increase their size with integration time. This effect has been observed in CCDs with laboratory data and on-sky measurements (\cite{m}), and has been interpreted as a consequence of electrostatic repulsion of incoming charge into neighboring pixels from charge previously accumulated in a pixel. Since some of the charge that would have been detected by a given pixel is being pushed into others, the counts in the pixels are not independent from each other and correlated, producing deviations from Poisson statistics. Smith (2008) (\cite{n}) proposed that these observed correlations could be interpreted in terms of effective pixel area variations. Antilogus et al. (2014) (\cite{l}) produce a phenomenological model for BF in CCDs in terms of these area variations (shifts of the pixels boundaries). The coefficients of the model can be obtained by measuring the correlations in flat-field images. This model was implemented and applied to data from the Dark Energy Survey (\cite{o}). 

\begin{figure}[htbp]
\centering 
\includegraphics[width=1.0\textwidth,origin=c]{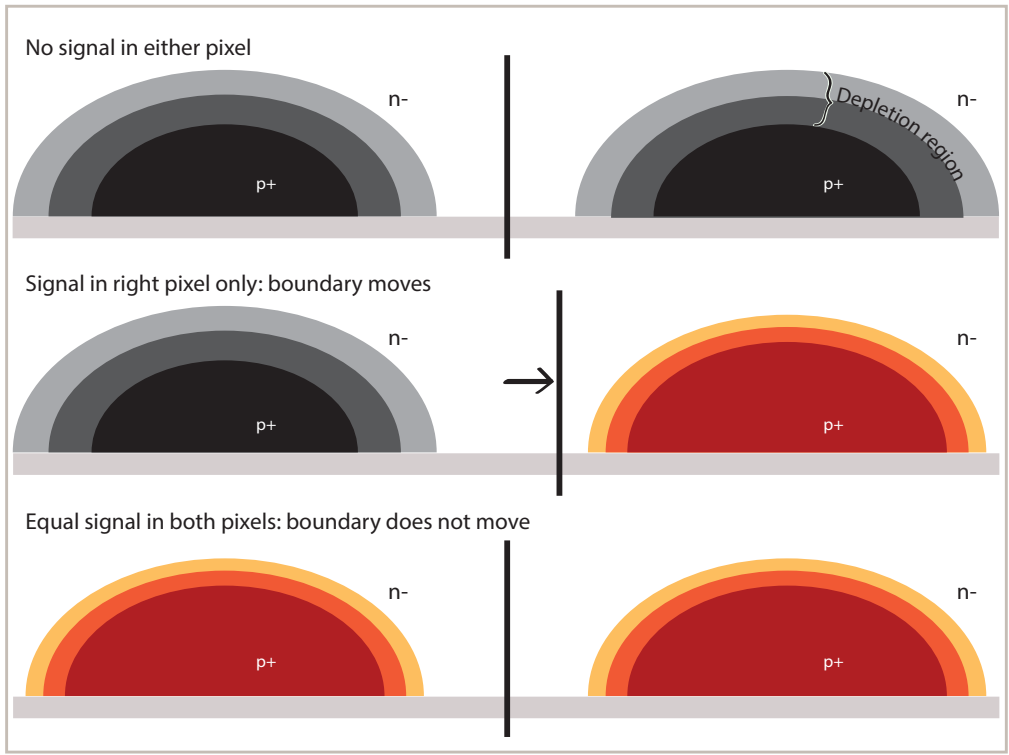}
\caption{\label{fig:bf_cartoon} Shift of the pixel boundary in NIR detectors as a consequence of the change in the size of the p-n junction depletion region due to the presence of charge (induced by illumination or the reset voltage). \emph{Upper panel}: In the absence of signal contrast (i.e., when there is no signal in either pixel or when the signal is the same) the pixel boundary lies midway between the depletion region edges. \emph{Middle panel}: When there is more signal in one of the two pixels (the one of the right in this case), the depletion regions of the pixel shrinks, causing the effective boundary between the pixels to move towards the pixel with higher signal. \emph{Lower panel}: If both pixels have the same signal, there is no boundary movement.}
\end{figure}

In NIR devices, the shrinking of the depletion region as photo-generated charge accumulates in a pixel provides a physical motivation for expecting a BF-type effect. Figure~\ref{fig:bf_cartoon} illustrates this. When there is no signal in two adjacent pixels, the boundary (which is not a physical barrier and, in the absence of charge diffusion, is defined as a plane that indicates by which pixel a charge will be captured) lies at the midpoint between the edges of the depletion regions. If charge starts to build-up in one of the pixels, its depletion region shrinks, moving the boundary towards the center of that pixel. The effect has more influence on those photons that land within the diffusion length of the boundary, as they will have more probability of being collected in the depletion region with greater area (the one with no charge in this case), effectively deflecting the charge. On the other hand, if both pixels have the same accumulated charge, the symmetry is restored and there is no boundary motion, demonstrating that this effect depends on the \emph{contrast} between pixels. 

\subsection{Analysis of WFC3-IR data}
\label{sec:wfc3_data}
The BF effect has been characterized in CCDs by measuring correlations in flat field images and by analyzing how the sizes (\emph{e.g.}, the half-light radii) of point sources increase with flux. With NIR detectors, it is possible to take advantage of the nondestructive reading per pixel and study the evolution of accumulated charge with time, as well as to analyze not only spatial but also temporal correlations.\footnote{When analyzing spatial correlations, the effect of IPC dominates, and must be taken into account.} 
\begin{figure}[htbp]
\centering 
\includegraphics[width=1.0\textwidth, page=1]{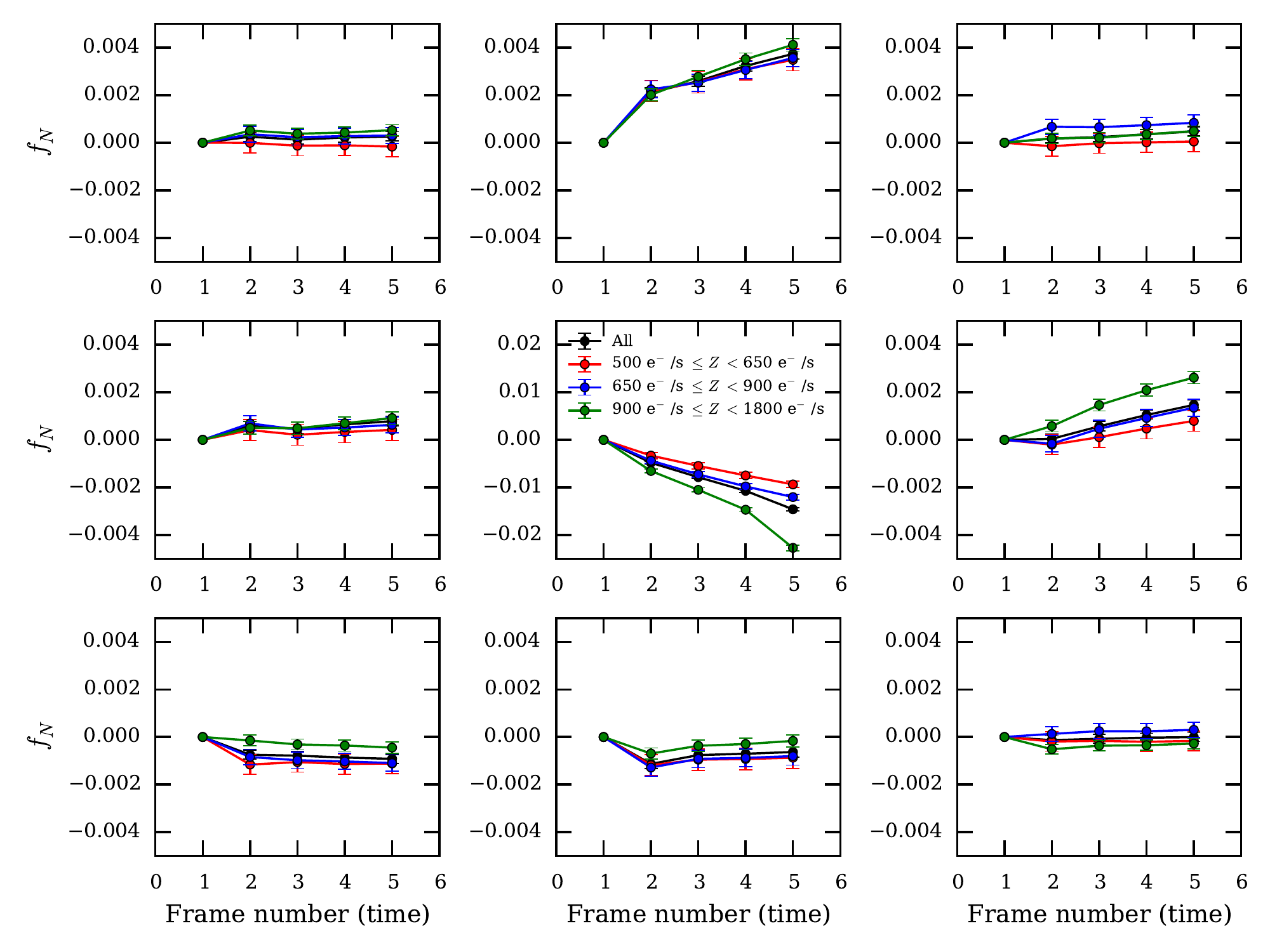}
\caption{\label{fig:wfc3_center} Averaged normalized difference between the flux at later reads and the flux at read $N=2$ the SUTR data from the WFC3-IR channel, as defined by Equation~\ref{eq:frac}. The signal has been divided in three $Z\equiv${\tt{FLUX\_AUTO}} (in units of e$^-$ per second) bins, represented by the red, blue, and green points. Each bin has roughly the same number of stars (1055, 996, and 1023, respectively) whose centroid is within $\pm$ 0.1 pixels from a pixel center, from a total of 3074 stars (black points). The error bars are given by the standard error of the mean. Each panel represents one pixel.}
\end{figure}
We use archival data from the WFC3-IR (an H1RG device with a pixel size of 18 $\mu$m) to look for the BF effect in stellar sources.\footnote{We follow the analysis by Jay Anderson (STScI), who presented preliminary results on the BF effect with this type of data (WFC3-IR channel) at a WFIRST meeting in May of 2016. Our analysis is an effort to validate those results and present them to a broader audience.} The data were downloaded from the Mikulski Archive for Space Telescopes database, and correspond to program 13606, which targets the globular cluster \emph{$\Omega$-cen}; it was taken about 6 arc minutes off the center of the cluster, providing about $\times$4 less crowding but still enough density of bright stellar sources. The processing of WFC3-IR exposures produces intermediate SUTR files with $N$ samples. For these images, $N=6$. These files (known as {\tt{MULTIACCUM}} or {\tt{IMA}} files) have reductions and calibrations applied, including dark frame subtractions, flat-fielding, and classical nonlinearity. For the latter, the correction is based on an average curve, fitted to a third degree polynomial (\cite{p}). IPC is not corrected, but it is approximately linear with signal and thus distinct from the BF effect. Each image is a multi-extension {\tt{FITS}} file with 6 samples, and each sample contains a science frame with the measured flux rate per pixel (electrons per second in this case), an error array, and a data quality map (to exclude saturated pixels, for example). 
\begin{figure}[htbp]
\centering 
\includegraphics[width=1.0\textwidth, page=2]{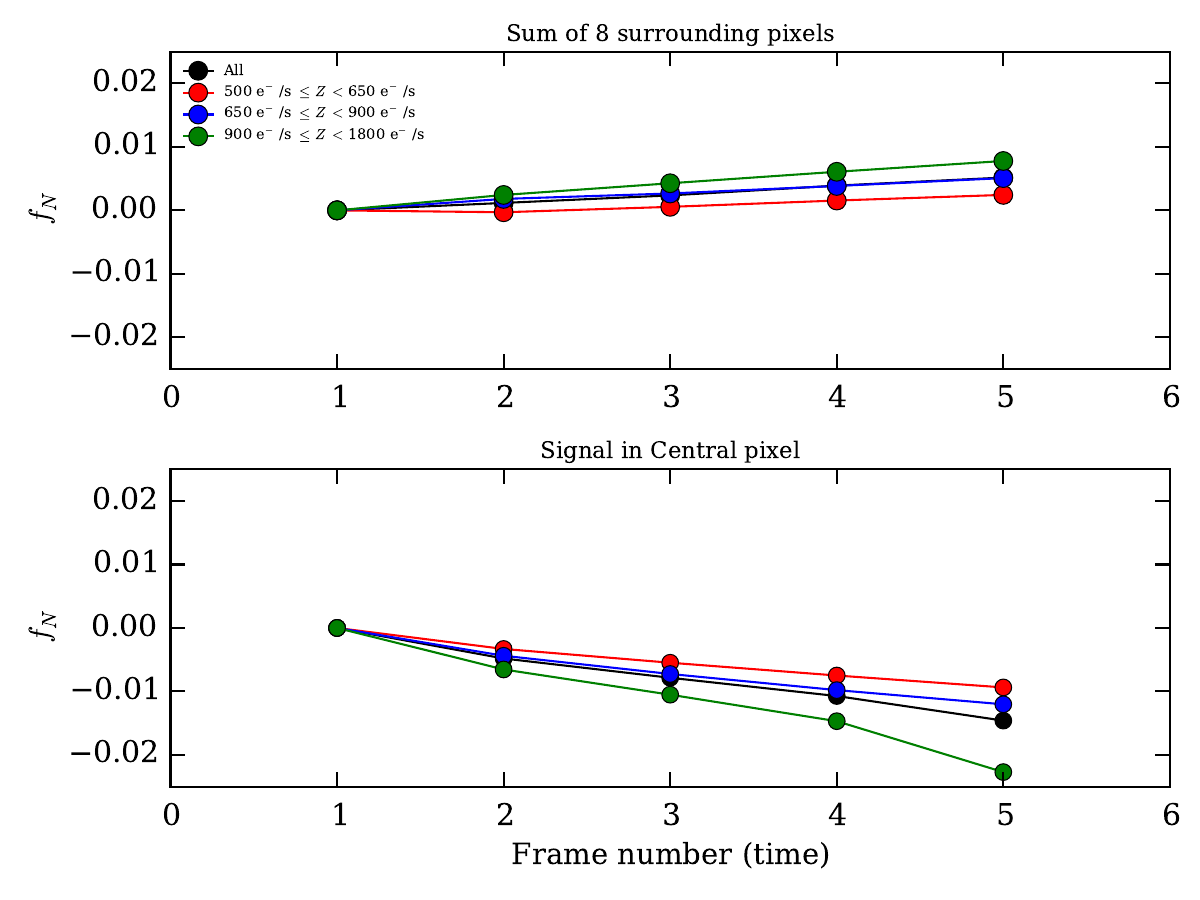}
\caption{\label{fig:wfc3_sum} \emph{Upper panel}: Sum of the 8 neighboring pixels that surround the central pixel in Figure~\ref{fig:wfc3_center}. \emph{Lower panel}: Central pixel from Figure~\ref{fig:wfc3_center}, reproduced here for comparison.}
\end{figure}
We use the software {\tt{SExtractor}} (\cite{q}) to detect stars in the sample with the longest exposure time for each image, using only the pixels flagged as ``good" by the data quality frame and {\tt{SExtractor}} (the flag is set to $0$ in both cases). We record the 3$\times$3 sub-array of pixels centered around the centroid of each star (as defined by the {\tt{SExtractor}} parameters {\tt{XWIN\_IMAGE}} and {\tt{YWIN\_IMAGE}}). To increase the contrast between pixels, we initially selected stars with centroid within $\pm 0.1$ pixels of a pixel center. Since these images have been reduced and corrected for NL, we should expect the ramp for each star to be linear, i.e. a constant flux rate (e$^-$ per second) between samples.
%we would expect each sample to posses the same flux rate (e$^-$ per second) and the ramp for each star to be horizontal. 
The BF effect would manifest as a decrease in flux at later reads compared to the earlier ones. For each star, we calculate the normalized difference:
\begin{equation}
\label{eq:frac}
f_{\text{N}}=\frac{F_{k} - F_{j}}{Z}, \ k\geq j
\end{equation}
where $F_{k,j}$ is the flux at reads $k$th and $j$th, respectively, with $k\geq j$. $Z$ is a normalization factor for each star and it is given in this case by the {\tt{SExtractor}} parameter {\tt{FLUX\_AUTO}}.\footnote{{\tt{FLUX\_AUTO}} is an adaptive aperture that performs elliptical aperture photometry within the Kron radius of the object.} We then stack these normalized differences (rejecting outliers with 3--$\sigma$ clipping) for each of the 9 pixels selected, and plot them as a function of sample number in $Z$ bins. 
\begin{figure}[htbp]
\centering % \begin{center}/\end{center} takes some additional vertical space
\includegraphics[width=1.0\textwidth, page=1]{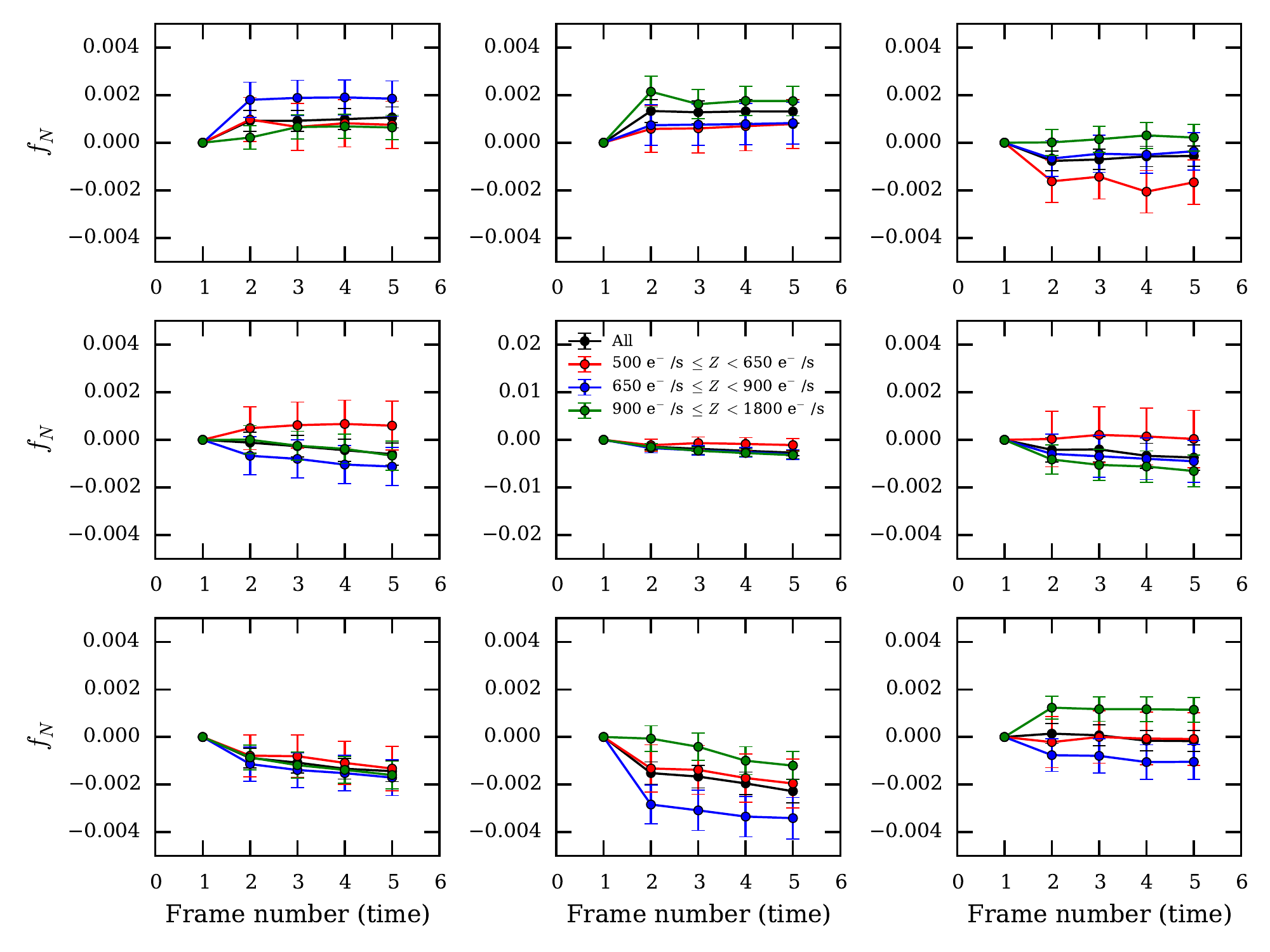}
\caption{\label{fig:wfc3_corner} Same as Figure~\ref{fig:wfc3_center}, but using stars whose centroid is within $\pm 0.1$ pixels of the corner of a given pixel. In this case, the red, blue, and green flux bins have 217, 216, and 214 stars, respectively, from a total of 647 (black).}
\end{figure}
The results are shown in Figure~\ref{fig:wfc3_center}, where each panel represent one pixel.  There is a general decrease in the average value of the central pixel with time, compared with what would be expected from the reference read at an earlier time (which in this case was taken to be the second read). The adjacent pixels seem to show an increase of flux with frame number, while the central pixel shows a steady decrease, indicating that some of the charge is being redistributed from the central pixel into its neighbors. When plotting the sum of the signal in the 8 pixels that surround the central pixel, a steady increase is observed instead (Figure~\ref{fig:wfc3_sum}). If the variation were due to pointing jitter or a similar effect, we would expect an increase in one side and a decrease in the other side. %In addition, the trends have been averaged over several stars. 
On the other hand, it could be that the NL correction applied by the reduction pipeline of the WFCE-IR channel still left some noticeable residuals. However, in this case, we would expect the adjacent pixels to have the same signal deficit as the central pixel. We also looked at stars whose centroids were within $\pm 0.1$ pixels of a pixel's corner (as opposed to the center), in which case the signal is greatly attenuated (Figure~\ref{fig:wfc3_corner}).   

\subsection{Prospects for laboratory characterization}
The results presented in Section~\ref{sec:wfc3_data} show preliminary evidence in favor of the presence of a BF effect in NIF detectors which could be interpreted in terms of shifting pixels boundaries. To confirm this, further analysis with other data sets is needed. In particular, laboratory analyses of point sources would provide great insight to characterize this effect. Laboratory tests have several advantages, such as uniformity in the spot spacing (avoiding source confusion), uniformity in the signal-to-noise ratio, and the ability to repeat measurements under controlled and stable conditions, among others. 

We are carrying out such tests at the PPL. Rapid data acquisition and tight control over stimuli and systematic errors make the PPL well-suited for investigating the BF effect and its dependence on the optical stimulus while validating the BF model. More analysis would be needed to assess its impact on WL science in the light of the requirements or the particular mission at hand, and devise a correction scheme. The phenomenological model proposed in Antilogus et al. (2014) and Guyonnet et al. (2015) (\cite{l,m})---which assumes that the displacements of the boundaries of a given pixel are linear in the charge stored in the surrounding pixels---as implemented by Gruen et al. (2015) (\cite{o}) could be used to perform such a correction. 

\section {Conclusion}
\label{sec:end}
Future space missions such as WFIRST will use a new family of NIR detectors to perform galaxy shape measurements for weak lensing to exquisite accuracy. These sensors have not been used before for this type of analysis, and like any other devices, they suffer from various effects that must be calibrated and characterized to satisfy the stringent requirements of future projects. In particular, we have discussed the intrinsic voltage nonlinearity due to the shrinking of the depletion region at the p-n junction as charge accumulates, and the deviation from linearity originating in the ROIC. We have presented laboratory measurements of this type of nonlinearity, and pointed out the need to carefully characterize it to calibrate it with a percent accuracy. In addition, using existing archival data from the WFC3-IR channel, we have presented evidence for the existence of a ``brighter-fatter"-type of effect in NIR devices, analogous to that measured in CCDs. We have proposed an interpretation of our analysis in terms of the shifting of pixel boundaries, and discussed future work to confirm and characterize this effect under the controlled conditions of a laboratory. 

\appendix
\label{ap:app}
\section{Quadratic approximation to classical nonlinearity}
It can be shown that the dependence on capacitance of voltage leads to an approximate quadratic response at low to medium signals. Inserting Equation~\ref{eq:cjn} in the standard relationship $V =Q/C$, we have:
\begin{equation}
V=\frac{Q}{A}\sqrt{1-\frac{V}{V_{\text{bi}}}}.
\end{equation}
Reorganizing the previous equation, we have:
\begin{equation}
\label{eq:a1}
\left( \frac{A}{Q} \right) ^2V^2 + \frac{V}{ V_{\text{bi}}} - 1 = 0.
\end{equation}
Equation~\ref{eq:a1} shows that V(Q) is, in principle, the positive root of a quadratic equation. The solution is:  
%To proceed, we use the standard solution to the quadratic equation $ax^2 + bx +c=0$, re-written as $x = b/(2a) ( -1 + (1 - 4ac/b^2)^{1/2} )$. Setting $a\equiv(A/Q)^2$, $b\equiv1/V_{\text{bi}}$, anc $c\equiv-1$, we obtain: 
\begin{equation}
\label{eq:ap2}
V=\frac{1}{2V_{\text{bi}}} \left ( \frac{Q}{A} \right )^2 \left ( -1 + \sqrt{1 + 4 \left ( \frac{A}{Q} \right)^2V_{\text{bi}}^2 } \right ).
\end{equation}
In the limit where $(2AV_{\text{bi}}/Q)^2$ << 1, and using the approximation $\sqrt{1+\delta x} \approx 1 + \delta x/2$ for $\delta x << 1$, Equation~\ref{eq:ap2} simplifies to $V\approx V_{\text{bi}}$, which is not relevant as this is the case when the p-n diode is forward-biased. On the other hand, when the diode is sufficiently reverse biased, $(2AV_{\text{bi}}/Q)^2$  >> 1 and Equation~\ref{eq:ap2}  becomes:
\begin{equation}
\label{eq:ap3}
V \approx \frac{Q}{A} -  \left( \frac{1}{2V_{\text{bi}}} \right ) \left ( \frac{Q}{A}\right )^2 .
\end{equation}
Equation~\ref{eq:ap3} is valid for sufficiently small signals but will break down as we approach saturation. Moreover, it could be used to estimate---to the extent of the assumptions made to derive it---the amplitude of classical nonlinearity  knowing the built in voltage and detector capacitance. 

\acknowledgments
We thank J. Anderson, C. Hirata, S. Deustua, J. Kruk, R. Mandelbaum, G. Bernstein, A. Kannawadi, D. Gruen, the Goddard Space Flight Center Detector Characterization Laboratory, and S. Adalsteinsson. We also thank the organizers of the workshop \emph{Precision Astronomy with Fully Depleted CCDs 2016}, especially A. Nomerotski. 

AAP is supported by the Jet Propulsion Laboratory. CS, JR, and EH are being supported in part by the Jet Propulsion Laboratory. The research was carried out at the Jet Propulsion Laboratory, California Institute of Technology, under a contract with the National Aeronautics and Space Administration.

\textcopyright 2017. All rights reserved.

\end{document}